\documentclass{sig-alternate-05-2015}
\newfont{\mycrnotice}{ptmr8t at 7pt}
\newfont{\myconfname}{ptmri8t at 7pt}
\let\crnotice\mycrnotice%
\let\confname\myconfname%

\usepackage{color}
\usepackage{colortbl}
\usepackage{pgf}
\usepackage{amssymb}
\usepackage{xspace}
\usepackage{tikz}
\usepackage{pgfplots}
\usepackage{footnote}
\usepackage{url}
\usepackage{multirow}
 \usepackage{todonotes}
 \usepackage[latin1]{inputenc}              
\usepackage[OT1]{fontenc}                  
\usepackage[english]{babel}                 
\usepackage{charter}                       
\usepackage{graphicx}
\usepackage{listings}
\usepackage{arydshln}
\usepackage{multicol}

\newcommand{\bip}{\ensuremath{\mathbb{B}}\xspace}
\newcommand{\bipedge}{\ensuremath{E_{\bip}}\xspace}

\newcommand{\set}[1]{\ensuremath{\{#1\}}}
\newcommand{\setst}[2]{\ensuremath{\{#1\ \vert\ #2\}}}

\CopyrightYear{2016} 
\setcopyright{acmcopyright}
\conferenceinfo{ICTIR '16,}{September 12-16, 2016, Newark, DE, USA}
\isbn{978-1-4503-4497-5/16/09}\acmPrice{\$15.00}
\doi{http://dx.doi.org/10.1145/2970398.2970413}

\begin{document}

\title{Exploiting the Bipartite Structure of Entity Grids for Document Coherence and Retrieval}
%
%
\numberofauthors{2} 
%
\author{
%
%
\alignauthor
Christina Lioma\\
       \affaddr{Department of Computer Science}\\
       \affaddr{University of Copenhagen, Denmark}\\
       \email{c.lioma@di.ku.dk}
\alignauthor
Fabien Tarissan\\
       \affaddr{ISP, ENS Cachan \& CNRS}\\
       \affaddr{University Paris-Saclay, France}\\
       \email{firstname.lastname@cnrs.fr}
\and
Jakob Grue Simonsen and Casper Petersen\\
       \affaddr{Department of Computer Science}\\
       \affaddr{University of Copenhagen, Denmark}\\
       \email{\{simonsen,cazz\}@di.ku.dk}
       \alignauthor
Birger Larsen\\
       \affaddr{Department of Communication}\\
       \affaddr{Aalborg University Copenhagen, Denmark}\\
       \email{birger@hum.aau.dk}
}
\date{30 July 1999}
%
\maketitle
\begin{abstract}
Document coherence describes how much sense text makes in terms of its logical organisation and discourse flow. Even though coherence is a relatively difficult notion to quantify precisely, it can be approximated automatically.  
This type of coherence modelling is not only interesting in itself, but also useful for a number of other text processing tasks, including Information Retrieval (IR), where adjusting the ranking of documents according to \textit{both} their relevance \textit{and} their coherence has been shown to increase retrieval effectiveness \cite{petersen,TanGP12}. 

The state of the art in unsupervised coherence modelling represents documents as bipartite graphs of sentences and discourse entities, and then projects these bipartite graphs into one--mode undirected graphs. However, one--mode projections may incur significant loss of the information present in the original bipartite structure. To address this we present three novel graph metrics that compute document coherence on the original bipartite graph of sentences and entities. 
Evaluation on standard settings shows that: (i) one of our coherence metrics beats the state of the art in terms of coherence accuracy; and (ii) all three of our coherence metrics improve retrieval effectiveness because, as closer analysis reveals, they capture aspects of document quality that go undetected by both keyword-based standard ranking and by spam filtering. 
This work contributes document coherence metrics that are theoretically principled, parameter-free, and useful to IR.

%
%
\end{abstract}

%
%

%

\section{Introduction}
\label{s:intro}
Document coherence is the logical organisation and development of thematic content in a document. The more coherent a document is, the more understandable it tends to be. Automatically measuring document coherence is useful for several tasks, such as text summarisation \cite{CelikyilmazH11,Parveen015,zhang:2011}, machine translation
\cite{LinLNK12,xiong:2013,Xiong:2015}, and information retrieval (IR) \cite{petersen,TanGP12}. For IR in particular, document coherence is typically treated as a feature of document quality (similarly to e.g.\ readability \cite{BenderskyCD11}, information-to-noise ratio \cite{ZhouC05}, or comprehensibility \cite{TanGP12}). Such document quality features have been found to improve retrieval performance when used to boost the ranking of documents which are more relevant \textit{and} of better quality. We present three new ways of measuring document coherence, and we practically show their usefulness to IR.

\begin{table*}[t]
\centering
\begin{minipage}{0.3\textwidth}
\centering
\setlength{\tabcolsep}{1pt}
\begin{tabular}{llllllllllllllll}
 & \rotatebox{90}{Department} & \rotatebox{90}{Trial} & \rotatebox{90}{Microsoft} & \rotatebox{90}{Evidence} & \rotatebox{90}{Competitors} & \rotatebox{90}{Markets}& \rotatebox{90}{Products} & \rotatebox{90}{Brands} & \rotatebox{90}{Case} & \rotatebox{90}{Netscape} & \rotatebox{90}{Software} & \rotatebox{90}{Tactics} & \rotatebox{90}{Government} & \rotatebox{90}{Suit} & \rotatebox{90}{Earnings}\\\hline
1 & \textbf{s}& \textbf{o}&\textbf{s} & \textbf{x}& \textbf{o}& --& --& --& --& --& --& --& --& --& --\\ 
2 & --& -- & \textbf{o} & --&-- & \textbf{x} & \textbf{s} & \textbf{o} & --& --& --& --& --& --& --\\
3 & --& --& \textbf{s} & \textbf{o} & --& --& --& --& \textbf{s} & \textbf{o}& \textbf{o}& --& --&-- &-- \\
4 & --&-- & \textbf{s} & --& --& --& --& --&-- & --& --&\textbf{s} & --& --& --\\
5 & --& --& --& --& --& --& --& --& --& --& --& --& \textbf{s} & \textbf{o}& --\\
6 & --& \textbf{x}& \textbf{s}& --& --& --& --& --& --& --& --& --& --& --& \textbf{o}\\
\end{tabular}
\end{minipage}%
\hfill
\begin{minipage}{0.7\textwidth}
\centering
\footnotesize
\begin{tabular}{lp{0.9\textwidth}}\noalign{\smallskip}
1 & [The Justice Department]$_{\mbox{\textbf{s}}}$ is conducting an [anti-trust trial]$_{\mbox{\textbf{o}}}$ against [Microsoft corp.]$_{\mbox{\textbf{s}}}$ with [evidence]$_{\mbox{\textbf{x}}}$ that [the company]$_{\mbox{\textbf{s}}}$ is increasingly attempting to crush [competitors]$_{\mbox{\textbf{o}}}$.\\ 
2 & [Microsoft]$_{\mbox{\textbf{o}}}$ is accused of trying to forcefully buy into [markets]$_{\mbox{\textbf{x}}}$ where [its own products]$_{\mbox{\textbf{s}}}$ are not competitive enough to unseat [established brands]$_{\mbox{\textbf{o}}}$.\\ 
3 & [The case]$_{\mbox{\textbf{s}}}$ revolves around [evidence]$_{\mbox{\textbf{o}}}$ of [Microsoft]$_{\mbox{\textbf{s}}}$ aggressively pressuring [Netscape]$_{\mbox{\textbf{o}}}$ into merging [browser software]$_{\mbox{\textbf{o}}}$.\\
4 & [Microsoft]$_{\mbox{\textbf{s}}}$ claims [its tactics]$_{\mbox{\textbf{s}}}$ are commonplace and good economically.\\
5 & [The government]$_{\mbox{\textbf{s}}}$ may file [a civil suit]$_{\mbox{\textbf{o}}}$ ruling that [conspiracy]$_{\mbox{\textbf{s}}}$ to curb [competition]$_{\mbox{\textbf{o}}}$ through [collusion]$_{\mbox{\textbf{x}}}$ is [a violation of the Sherman Act]$_{\mbox{\textbf{o}}}$.\\
6 & [Microsoft]$_{\mbox{\textbf{s}}}$ continues to show [increased earnings]$_{\mbox{\textbf{o}}}$ despite [the trial]$_{\mbox{\textbf{x}}}$.\\
\end{tabular}
\end{minipage}
\label{tab:grid}
 \caption{Entity grid example from \cite{Barzilay:2008}. Discourse entities are inside square brackets, marked $s,o,x$ for subject, object and other grammatical role, respectively.} 
\end{table*}

The starting point of our work is the linguistic definition \cite{BeaugrandeD81} of document coherence as the thematic unity that stems from the links among the underlying ideas of a document and from the logical organisation and development of its content. It is this \textit{logical continuity of senses} that characterises coherent documents and makes them overall understandable. This continuity of senses is traditionally modelled as the transition of topics throughout sentences. 
Typically this topic transition is approximated by extracting salient discourse entities from a document (for instance, the subject and object of each sentence) and measuring their occurrence (and distance) through sentences in an \textit{entity grid} \cite{Barzilay:2008} (see example in Table \ref{tab:grid}). Recently the elements of such an entity grid (i.e., the discourse entities and the sentences in which they occur) have been represented as a graph, the topology of which has been used to approximate document coherence, for instance as the average out-degree \cite{strube}, pagerank, clustering coefficient, or betweenness \cite{petersen} computed over the whole graph (each graph representing a single document). 
This type of graph-based coherence modelling, despite being completely unsupervised, performs comparably to equivalent supervised approaches, thus showing great promise. We posit that these existing graph-based computations of document coherence are suboptimal in capturing the transition of entities across sentences, and we present a principled solution for improving this. We explain this next.

\begin{figure}[]
\centering
\scalebox{0.5}{
\includegraphics[scale=1.0]{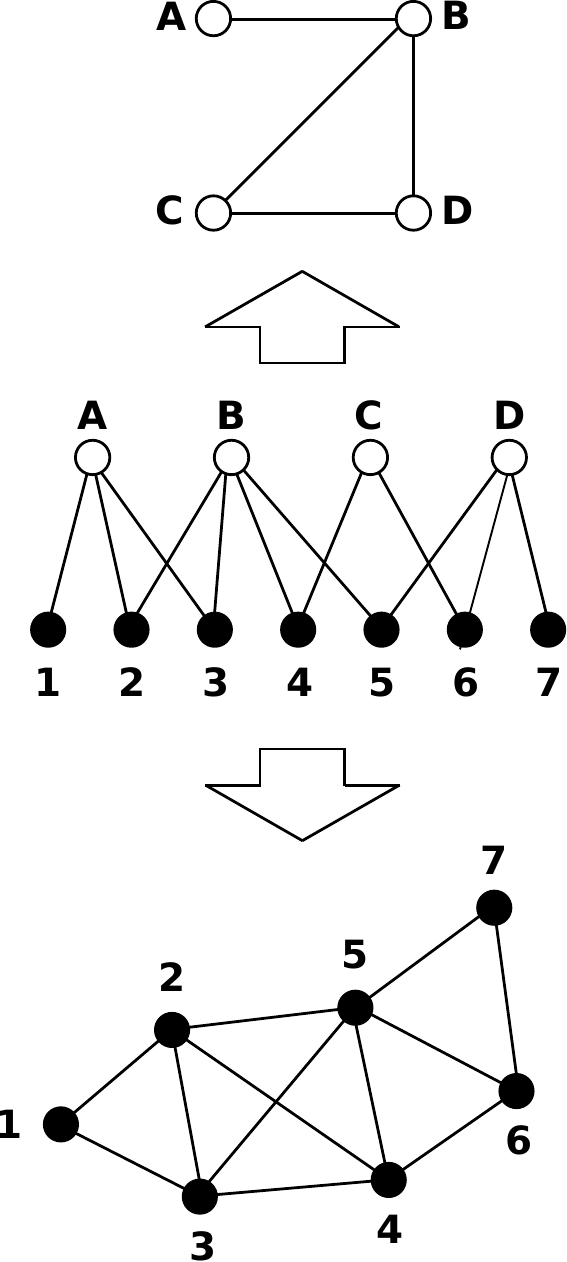}
}
\caption{Bipartite graph (middle) and its two projections (top and bottom), from \cite{Newman}. A--D and 1--7 denote two different types of vertices.}
\label{fig:projection}
\end{figure}

Existing graph-based computations of text coherence \cite{strube,petersen} represent each document as a \textit{bipartite graph} of sentences and their discourse entities. A bipartite graph is a particular class of graph also known as \textit{two-mode} graph. 
In the case of coherence, one type of vertices denotes discourse entities, and the other type denotes the sentences in which these entities appear. The edges in a bipartite graph in typical coherence modelling connect only vertices of unlike types, i.e. only entities and sentences. Current graph-based coherence approaches project this bipartite graph of entities and their sentences onto a one-mode graph of only sentences that are connected if they contain at least one common entity. This projection is motivated by convenience: it is easier to work with direct connections between vertices of just one type. All current graph-based metrics of document coherence are computed solely on such one-mode projections of bipartite graphs. 

Even though one-mode projections of two-mode graphs are widely employed, they are a less powerful representation of the data because they discard part of the information present in the structure of the original bipartite graph. 
This point is graphically illustrated in Figure \ref{fig:projection}, which shows a bipartite graph (middle) and its two one-mode projections (top and bottom). For entity bipartite graphs, the analogy would be that the sentences of a document are denoted by A-D and its entities are denoted by 1-7. As Figure \ref{fig:projection} shows, part of the information captured by the bipartite graph about the entity transition throughout the document is lost or compressed with one-mode projections. We reason that this information loss is propagated to any coherence metric that is then computed on an one-mode projection of such a graph, resulting in potentially suboptimal approximations of document coherence. 

We present three coherence metrics that are applied directly on the original bipartite graph, not on its one-mode projection (Section \ref{s:model}). 
Our metrics are new, constituting a contribution not only to coherence modelling but also to graph metrics.
In addition, our bipartite metrics incur no additional efficiency cost over existing one--mode graph metrics. One of our metrics is shown to be a \textit{much} more accurate approximation of document coherence than the state of the art computed from one-mode projections \cite{strube,petersen} (Section \ref{s:results-coh}). All three coherence metrics are shown to be useful to retrieval effectiveness (Section \ref{s:results-ir}). To our knowledge, such two-mode graph-based coherence metrics have not been investigated before.


\section{Related Work}
\label{s:relw}

Several metrics using the entity grid (or extensions thereof) have been proposed for approximating the coherence of a
document (see \cite{petersen,Xiong:2015,ZhangFQHLH15} for recent overviews). Broadly these methods compute probabilities of entity transitions on the grid, and use these probabilities to learn coherence in a supervised way. The particular line of research extending the entity grid that is relevant to our work transforms the entity grid into a bipartite graph of sentences and entities~\cite{strube}. Coherence is then approximated in an unsupervised way as the average out-degree~\cite{strube}, pagerank, clustering coefficient, betweeness centrality, entity distance, or adjacent and non-adjacent topic flow~\cite{petersen} on one-mode projections on the sentence vertices of that graph (equivalent to the top projection in Figure \ref{fig:projection}). 
This process of reducing a two-mode ``entity--and--sentence'' graph into an one-mode ``sentence--only'' graph loses all information about \textit{how many} and \textit{which} entities two sentences share, as well as the \textit{exact} entities occurring in a given sentence. Part of this information can be captured in one-mode projections by making the projection weighted. This has been done \cite{strube}, by weighting each edge in the projection by the number of entities its two connecting vertices share. This type of weighted projection retains information about how many entities two sentences share, but still fails to capture the identity and the transition of those entities across sentences, thus removing the option of drawing entity-oriented insights from the graph. Another interesting weighted one-mode projection of such bipartite graphs has also been presented \cite{strube}, which weights edges according to the grammatical roles of the entity vertices they share. This has been done by assigning arbitrary scores of 3, 2, 1 for the grammatical roles of subject, object or other, respectively, and then summing these scores over all shared entity vertices between two sentence vertices. This projection, despite being weighted, does not compensate for any information loss incurred by compressing a bipartite graph into an one-mode projection, but rather it attempts to enrich the graph with grammatical information. To our knowledge, our work is the first to propose coherence metrics computed directly on the bipartite graph and not on its one mode projections. 

The document coherence metrics we present can be seen as estimating an aspect of \textit{document quality}. 
A wide variety of document quality aspects have been used in IR, ranging from heuristics on document format (e.g., the fraction of anchor text on a hyperlinked document \cite{NtoulasNMF06}), to hyperlinked-derived estimations of popularity (e.g., PageRank, HITS \cite{NtoulasNMF06}). Another common type of document quality approximations are content-based. These are numerous and diverse, including for instance, ratios of 
information-to-noise, of stopwords per document, or of document words per stopword list \cite{BenderskyCD11, ZhouC05,ZhuG00}; 
average term length per document \cite{KanungoO09}; term part-of-speech \cite{LiomaO07, LiomaK:2008};  
ratio of technical terminology per (scientific) document \cite{LarsenL12}; ratio of non-compositional phrases per document \cite{MichelbacherKFLS11}; syllable, term and/or sentence statistics \cite{TanGP12} as per standard readability indices \cite{ColemanL75, Gunning52, KincaidFRC75, McClure87, McLaughlin69}; discourse structure \cite{Lioma12}; document entropy computed from terms \cite{BenderskyCD11} or discourse entities \cite{petersen}.
The lexical or syntactic features used in the above content-based document quality approximations are assumed to indicate syntactic or semantic difficulty. They are thus used to compute scores of document quality aspects such as readability, cohesiveness, comprehensibility or coherence, which are generally found to improve retrieval effectiveness when integrated into ranking, in particular with respect to precision at ranks 1--20 \cite{BenderskyCD11,petersen,TanGP12}.

In addition to using document coherence for improving IR, the reverse has also been reported, namely using IR to improve coherence modelling \cite{ZhangFQHLH15}. The idea here is to link entities that have different lexical form but are semantically related (e.g. \texttt{Gates} and \texttt{Microsoft}), by retrieving mentions of those entities from multiple web sources and mining their relations. This approach gives good performance. Interestingly, when mining such relations between entities from web data, the task of characterising the type of these relations has also been addressed using graph representations and has been modelled as an IR, and specifically learning to rank, problem \cite{VoskaridesMTRW15}.

To our knowledge, the current state of the art in coherence modelling in terms of accuracy is the deep learning approach of \cite{LiH14a}, where a recursive neural network learns sentential compositionality and is then used to model document coherence. This approach is supervised and computationally much heavier than graph-based coherence modelling.

\section{Bipartite graph metrics of \\document coherence without projection}
\label{s:model}
\def\fabien#1{\todo[inline, color=blue!40]{#1}}

Our work builds on the early assumption~\cite{grosz:1995} that a document is more coherent if its adjacent or near-adjacent sentences refer to the same entities. This \textit{transition} of entities across sentences is typically represented as an \textit{entity grid} \cite{Barzilay:2008}. The entity grid of a document is defined as a table whose rows
represent (consecutive) sentences in that document, and whose columns represent discourse entities that occur in that document. Each cell $(i, j)$ is either empty or contains
information about the syntactic, discourse, or other grammatical role of entity $j$ in sentence $i$. Table~\ref{tab:grid} displays an example of an entity grid borrowed from \shortcite{Barzilay:2008}. 
%

\begin{figure}[t!]
\begin{center}
\resizebox{\columnwidth}{!}{
    \begin{tikzpicture}[node distance=40pt, very thick]
   \node[rectangle,draw,](A) at (-1,0) {$S_1$};
    \node[rectangle, right of=A, draw] (B) {$S_2$};
    \node[rectangle, right of=B, draw] (C) {$S_3$};
    \node[rectangle, right of=C, draw] (D) {$S_4$};
    \node[rectangle, right of=D, draw] (E) {$S_5$};
    \node[circle,draw](3) at (0.5,-2) {$e_3$};
    \node[circle, left of=3, draw] (2) {$e_2$};
    \node[circle, left of=2, draw] (1) {$e_1$};
    \node[circle, right of=3, draw] (4) {$e_4$};
    \node[circle, right of=4, draw] (5) {$e_5$};
    \node[circle, right of=5, draw] (6) {$e_6$};
    \node[circle, right of=6, draw] (7) {$e_7$};
    \draw[-] (1) to  (A);
    \draw[-] (2) to  (A);
    \draw[-] (2) to  (B);
    \draw[-] (3) to  (B);
    \draw[-] (4) to  (B);
    \draw[-] (5) to  (B);
    \draw[-] (3) to  (C);
    \draw[-] (4) to  (C);
    \draw[-] (5) to  (C);
    \draw[-] (3) to  (D);
    \draw[-] (4) to  (D);
    \draw[-] (5) to  (D);
    \draw[-] (6) to  (D);
    \draw[-] (7) to  (D);
    \draw[-] (6) to  (E);
    \draw[-] (7) to  (E);
  \end{tikzpicture}
}  
\end{center}
\caption{An example of a bipartite graph of an entity grid. Sentences are represented by squares, entities by circles. Syntactic roles are omitted for readability.}
\label{bip}
\end{figure}
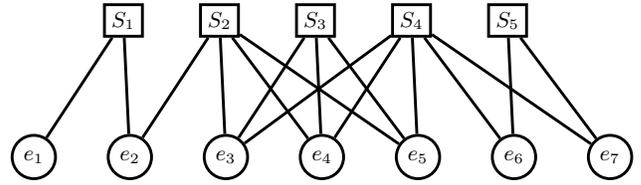

Following \cite{strube}, we represent the entity grid as a bipartite graph $\bip=(V_\top,V_\bot,\bipedge)$, where $V_\top$ is the set of \emph{sentences} in the document,
$V_\bot$ is the set of \emph{entities} in the document, and $\bipedge\subset \top\times\bot$ is the
set of edges relating entities to the sentences in which the entities
appear, each edge labelled with the value in cell $(i, j)$. An example of such a bipartite graph \bip is given in Figure~\ref{bip}, where
$\top$ vertices (sentences) are depicted by squares
($V_\top=\set{S_1,S_2,S_3,S_4,S_5}$) and $\bot$ vertices (entities) are depicted by
circles ($V_\bot=\set{e_1,e_2,e_3,e_4,e_5,e_6,e_7}$).
Every such bipartite graph can be \emph{one--mode projected},
resulting for instance in a graph $G = (V,E)$ where $V = V_\top$ and
$E = \setst{(u,v)\in V^2}{\exists w\in V_\bot, (u,w)\in\bipedge \textrm{ and }
   (v,w)\in \bipedge}$~\footnote{\scriptsize{A dual projection can be defined for
  $\bot$ nodes.}} (see example in Figure \ref{fig:projection}). This
projection allows to re-use all metrics defined for one-mode graphs,
but we claim that valuable information is discarded in the process.

In this paper, we extract information about coherence
\emph{from the bipartite structure itself}, without projecting the
structure over entities nor sentences. We reason that principles from
existing research on using (projected) graphs for coherence can be
retained, and that bespoke metrics for bipartite graphs can be
combined with these principles. We identify two primary
such principles:
\begin{description}
\item[Length of paths:] short paths between vertices representing
  entities generally imply rapid cognitive information processing~\cite{lioma}, hence are good indicators of coherence.
\item[Local density:] locally dense documents tend to
  involve successive sentences that share similar
  concepts and entities, hence are good indicators of coherence. 
\end{description}

In the remainder of this section, we propose coherence metrics that align with the above principles and are thus 
able to capture such short path or local density properties in the bipartite structure.  It
is worth noting that even though the notion of local density is well known for
one-mode graphs, there exist no standard definitions for it for bipartite
graphs. Indeed, the local density (usually captured by the
\emph{clustering coefficient} and the \emph{transitive ratio})
consists in computing the number of triangles (three vertices, all
connected) present in the graphs but, by definition, no such pattern
can exist in a bipartite graph. 

However, several extensions of the clustering coefficient have been
proposed~\cite{borgatti1997network,socnet,opsahl2013triadic,snijders2001statistical,zhang2008clustering}
to serve as proxy for this notion in bipartite graphs. These proxies have proven
to be useful in many contexts, ranging from improving the modelling of
the large-scale link structure of the
Internet~\cite{tarissan2013inet}, to analysing online social
networks~\cite{tarissan2015socnet}, or detecting landmark decisions in
judicial decision networks~\cite{tarissan2016icc}.

Next, we show how to adapt those definitions to
account for the specific context of local coherence estimation. 
We do so, reasoning on a bipartite graph, where $u
\in V_\top$, and where we define $N_{\top}(u) = \setst{v \in \bot}{(u,v)
  \in \bipedge}$ as the subset consisting of the vertices in $V_\bot$
that are linked to $u$.

\subsection{Bipartite distance-based clustering coefficient ($\mathtt{bipDCC}$)}
We call our first coherence metric \textit{bipartite distance-based clustering coefficient} ($\mathtt{bipDCC}$). The \emph{clustering
  coefficient} in standard graphs quantifies, roughly, how dense the
  graph is around its vertices. In our case, we are interested in estimating the extent to which successive sentences share similar or identical\footnote{\scriptsize{The original definition of the entity grid allows to model the succession of only identical entities. This is the definition we adopt here. However, our coherence metrics also work for extensions of the entity grid that capture similar but not identical entities \cite{ZhangFQHLH15}.}} entities (suggesting coherence). To do so, we propose the following adaptation of the bipartite clustering coefficient~\cite{socnet}.
Given two sentences $s_i$ and $s_j$ that have at least one entity in common:
\begin{enumerate}
\item Compute the fraction of shared entities with respect to the number of total entities occurring in $s_i$ and $s_j$ (classic notion of bipartite clustering coefficient based on the Jaccard index~\cite{socnet}), and
\item Account for the relative position of the involved sentences by dividing the former quantity by the distance between $s_i$ and $s_j$.

\end{enumerate}
Formally, assuming that $i$ and $j$ denote the position of sentences $s_i$ and $s_j$ in the document:
\begin{equation}
\mathtt{bipDCC}_{\top}(s_i,s_j) = \frac{1}{|j-i|} \cdot \frac{|N_{\top}(s_i)\cap N_{\top}(s_j)|}{|N_{\top}(s_i) \cup N_{\top}(s_j)|}
\end{equation}
For instance, in Figure~\ref{bip},  $\mathtt{bipDCC}_{\top}(s_i,s_j)$ would attain its highest values for vertices $S_2$ and $S_3$ ($\mathtt{bipDCC}(S_2,S_3)=0.75$).

Then, we define $\mathtt{bipDCC}_{\top}(s_i)$ as the average value of \\
$\mathtt{bipDCC}_{\top}(s_i,s_j)$ for all $s_j$ that share at least one
entity with $s_i$. We compute the bipartite distance-based clustering coefficient,
$\mathtt{bipDCC}_{\top}(\bip)$, of the entire bipartite graph of the
document as the average value of $\mathtt{bipDCC}_{\top}(s_i)$ for all
sentences $s_i$.

The intuition is that a coherent document will involve successive (or
almost successive) sentences sharing similar or identical entities, thus increasing
the value of the bipartite distance-based clustering coefficient.

Regarding the complexity of Equation 1, we note that by properly implementing the set operations (e.g.,
as bitwise operations on boolean strings) the worst-case complexity of
computing the right-hand side of the formula is linear in the total number of
bottom nodes.

\subsection{Bipartite asymmetric clustering coefficient ($\mathtt{bipACC}$)}

The bipartite distance-based clustering coefficient proposed above gives a similar
role to $s_i$ and $s_j$. In particular, it does not account for the
number of entities related to each of the sentences. This raises some
issues for small (in terms of number of entities) sentences. In Figure~\ref{bip} for instance, the reader might notice that
$\mathtt{bipDCC}_{\top}(s_5) = 0.4$, although the only two entities
involved in $s_5$ are both shared with another sentence, which turns
out to be the closest in the document. As such, the coefficient should
be the highest value ($1.0$).


In order to account for this, we propose the following
\emph{asymmetric} variant of $\mathtt{bipDCC}$.
Given two sentences $s_i$ and $s_j$ that share at least one common entity:
\begin{enumerate}
\item Compute the fraction of shared entities with respect to the number of entities that \emph{$s_i$ could have shared} with $s_j$, and
\item Use this fraction to discount the distance between $s_i$ and $s_j$, as previously for the bipartite distance-based clustering coefficient.
\end{enumerate}

Formally, we define the bipartite asymmetric clustering coefficient of $s_i$ and $s_j$ as:
\begin{equation}
\mathtt{bipACC}_{\top}(s_i,s_j) = \frac{1}{|j-i|} \cdot \frac{|N_{\top}(s_i)\cap N_{\top}(s_j)|}{|N_{\top}(s_i)|}
\end{equation}

Then, the bipartite asymmetric clustering coefficients of vertex $s_i$
($\mathtt{bipACC}_{\top}(s_i)$) and of the whole document ($\mathtt{bipACC}_{\top}(\bip)$) are respectively derived as averages, in the same way as for the distance-based clustering coefficient presented above. Note that while $\mathtt{bipDCC}_{\top}(s_i,s_j) = \mathtt{bipDCC}_{\top}(s_j,s_i)$, we have in general $\mathtt{bipACC}_{\top}(s_i,s_j) \neq \mathtt{bipACC}_{\top}(s_j,s_i)$ now.

By using this asymmetric variant of the bipartite distance-based clustering
coefficient, we expect to highlight in particular short sentences that
are well connected to each other. This might be particularly useful in domains where this type of writing is predominant (although we do not evaluate this potential domain adaptivity of this coherence metric in this work).

\subsection{Bipartite Linkage Coefficient ($\mathtt{bipLC}$)}

The two coefficients proposed so far are straight-forward variants of the original
bipartite clustering coefficient that attempt to capture local
density in bipartite graphs. However, it has been shown
in~\cite{socnet} that such coefficients might miss some important
properties of the overlapping between $\top$ vertices (in our case sentence vertices) in the bipartite
structures. This is why~\cite{socnet} suggested to use the
\emph{redundancy coefficient} $\mathtt{rd_\top}(v)$ of a vertex. The redundancy coefficient focuses on the impact of removing $v$ in regards to the
$\bot$-projection. 


To illustrate this impact on the example of Figure~\ref{bip}, consider sentences $S_1$ and $S_5$. Although they are both related to
two entities, they have a very different
way to relate to the rest of the sentences. One way to measure this
consists in projecting the bipartite graph over the entities and
comparing the resulting structure to the same projection if we remove
$S_1$ or $S_5$. Removing vertex $S_1$ results in
the loss of one edge (between $e_1$ and $e_2$). In contrast, if we
look at the impact of removing vertex $S_5$, the projection is exactly the same with or
without the vertex because the two entities it relates ($e_6$ and
$e_7$) are also related by sentence $S_4$. In this respect, $S_5$ is
said to be \emph{redundant}. The above are two extreme cases; in practice a wide range of situations usually depict different levels of
redundancy.


Following this principle of graph redundancy, and letting $v \in V_\top$, we define $D_v$ as the set 
{\small
$$D_v = | \set{\set{u,w}\in N_\top(v)^2 \vert \exists v'\not= v, (v',u)\in \bipedge \textrm{ and } (v',w)\in \bipedge}|$$
}.

\noindent That is, $D_v$ is the set of pairs of entities in sentence $v$ such that there is (at least) another sentence containing both of them. The
\emph{redundancy} of a node $s \in V_{\top}$ is then formally defined
as:
\begin{equation}
\mathtt{rd_\top}(v)=\frac{D}{\frac{|N_{\top}(v)| (|N_{\top}(v)|-1)}{2}}
\end{equation}

Intuitively, a high value of the $\mathtt{rd_\top}(v)$ indicates that
two entities that $v$ relates are likely to be related by another
sentence.  In the example above, $\mathtt{rd_\top}(v)$ assumes its
highest values for sentences $S_3$ and $S_5$. This is expected because
all entities in these sentences occur in (perhaps several) other
sentences.


As we wish to model coherence, and there is a natural order on sentences, we define a new variation of the redundancy that also captures closeness, which we call \textit{bipartite linkage coefficient} ($\mathtt{bipLC}$) as follows. Given a sentence $s_i$:
\begin{enumerate}
\item For each pair of entities $(e_k, e_l)$  in $s_i$, compute
  the distance between $s_i$ and the closest sentence that contains also
  $e_k$ and $e_l$ ($\infty$ if there is no such other sentence), and
\item Compute the average of the inverse of the distances computed in step 1.
\end{enumerate}
Formally, $\forall s_i$ and $\forall e_k, e_l \in N_\top(s_i)$, let $d_{ikl} = \min\setst{|j-i|}{s_j\in N_\bot(e_k) \cap N_\bot(e_l) - \{s_i\}}$. We define:
$$
\mathtt{d_{s_i}}(e_k,e_l)=\left\{
\begin{array}{ll}
\infty  & \textrm{ if } N_\bot(e_k) \cap N_\bot(e_l) = \set{s} \\
d_{ikl} & \textrm{ otherwise} 
\end{array}
\right.
$$

Then, the \emph{bipartite linkage coefficient} of a sentence $s_i$ is:
\begin{equation}
\mathtt{bipLC_\top}(s_i)=\frac{\sum_{e_k, e_l\in N_\top(s_i)} \frac{1}{\mathtt{d_{s_i}}(e_k,e_l)}}{\frac{|N_{\top}(s_i)| (|N_{\top}(s_i)|-1)}{2}}
\end{equation}

We then compute the linkage coefficient of the entire document $\mathtt{bipLC}_{\top}(\bip)$ as the average value of
$\mathtt{bipLC}_{\top}(s_i)$ for all sentences $s_i$. 

This coefficient is interesting because it 
explicitly relates the property of the
bipartite structure to the one of the $\bot$-projection, i.e. to the projection of entities, which are central in modelling coherence.

\section{Coherence Evaluation}
\label{s:results-coh}
Before evaluating the effectiveness of our coherence metrics for IR, we perform a pre-study to assess how accurately our coherence metrics approximate actual coherence. 

\subsection{Experimental Setup}
We use the standard dataset for coherence evaluation, \emph{Earthquakes and Accidents}\footnote{\scriptsize{\url{http://people.csail.mit.edu/regina/coherence/}}}, which contains 200 newswire articles (henceforth documents) concerning earthquakes and accidents from the North American News Corpus and 
 the National Transportation Safety Board. These documents are short (240 terms on average); we expect them to be coherent because they have been produced by human professionals aiming to inform the public.
We parse these documents with the Stanford parser and consider as entities those words tagged by the parser as the subject(s) or object(s) of a sentence. We do not treat as entities words of other grammatical roles (marked $x$ in Table \ref{tab:grid}) because we wish to consider only the most salient entities (i.e. the closest approximation to topics) of a document, and not modifiers of those topics by e.g. prepositional or other peripheral phrases (such as \texttt{with evidence, through collusion, into markets} in Table \ref{tab:grid}). We use the extracted entities to build entity grids, represent them as bipartite graphs, and compute our three coherence metrics as described in Section \ref{s:model}. All three of our coherence metrics are unsupervised -- they contain no parameters, hence no training is involved. 

We compare our coherence metrics to four coherence modelling baselines: 
\begin{enumerate}
\item Barzilay and Lapata's seminal entity grid model \cite{Barzilay:2008},
\item  Barzilay and Lee's HMM-based model \cite{BarzilayL04}, 
\item Guinaudeau and Strube's out-degree graph-based coherence metric \cite{strube}, and 
\item Petersen et al.'s entity distance graph-based coherence model (which is the best performing of all 11 coherence models presented in \cite{petersen}). 
\end{enumerate}
Note that baselines (1) and (2) are not graph-based,  and that baselines (3) and (4) use undirected one-mode projections; on the contrary, our $\mathtt{bipDCC}$, $\mathtt{bipACC}$, and $\mathtt{bipLC}$ coherence metrics are defined (and computed) directly on the bipartite graph of a document's entity grid, rather than on its one-mode projection.

We use the standard practice of evaluating coherence, which consists of
re-ordering progressively larger numbers of sentences in actual, coherent documents. This has the effect of simulating grades of incoherence; hence, good coherence metrics would have \emph{high} coherence scores for the original documents, but progressively \emph{lower} coherence scores as more and more sentences are re-ordered. 

 For each document, we 
pick $n \in [1 \; .. \; 20]$ pairs of sentences at random and switch them (e.g., for $n = 20$, a total of $40$ sentences switch places).
We then compute our coherence metrics on both the original and each of the re-ordered documents. 
If the coherence score of the original document is not lower than the coherence score of the re-ordered document, then we reason that the coherence metric accurately predicts the re-ordered document to be less coherent that the original. The total number of accurate predictions is then averaged over all documents.


%


\begin{table}[!ht]
\centering
\begin{tabular}{lccc}
\hline

                                             & Earthquakes &Accidents & Average\\
                                             \hline
Entity-grid      & 69.7    & 67.0 & 68.4\\
(no graph) 	&&&\\ \hdashline
HMM               & 60.3         & 31.7 & 46.0\\
(no graph) 	&&&\\ \hdashline
Out-degree 		&78.0	&80.0	&79.0\\
(one-mode projection) 	&&&\\ \hdashline
Entity distance 	  & 76.0      & 75.0	&75.5\\
(one-mode projection) 	&&&\\
\hline                                             
$\mathtt{bipDCC}$             & 55.6     & 69.8  & 62.7\\    
(bipartite graph)	&&&\\ \hdashline
$\mathtt{bipACC}$             & 55.5     & 70.1  & 62.8\\ 
(bipartite graph) 	&&&\\ \hdashline
$\mathtt{bipLC}$               & \bf 80.9     &\bf 94.0  &\bf 87.5\\
(bipartite graph)	&&&\\ 
\hline
\end{tabular}
\caption{Average coherence accuracy of baselines (top 4 rows) and our metrics (bottom 3 rows). The highest score in each column is shown in \textbf{bold}.}
\label{tab:coh}
\end{table}

\subsection{Coherence Accuracy Findings}
Table \ref{tab:coh} shows the accuracy of each coherence metric averaged over all re-ordered documents. We see that our $\mathtt{bipLC}$ metric is the most accurate, both on the individual subsets of the data, and on their overall average. Our two other metrics ($\mathtt{bipDCC, bipACC}$) are less accurate than $\mathtt{bipLC}$ but also than the two graph-based baselines computed on one-mode projections (out-degree and entity distance) and the original entity grid, on average. A possible explanation is that $\mathtt{bipDCC}$ and $\mathtt{bipACC}$ focus primarily on local clusters of entities, whereas the other metrics focus more on entity linkages across sentences (in different ways for each metric). This emphasis on entity linkage as opposed to entity clustering is possibly better suited to approximating coherence, as all currently best performing graph-based models of coherence \cite{strube, petersen} -- which we use as baselines (3) and (4) -- prioritise the
(graph-based) distance between entities in a document. In particular, the best performing $\mathtt{bipLC}$ metric emphasises the likelihood that two entities that are linked by a sentence will be linked by another sentence too. This property is related to the theory of lexical chains \cite{HallidayH76}, an early foundation in coherence modelling. To the best of our knowledge, of all existing coherence metrics, our $\mathtt{bipLC}$ metric models this idea of taking into account the trajectory of an entity across sentences the closest.

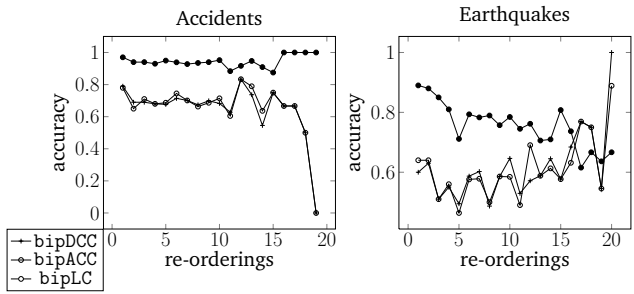
\begin{figure}
\centering
\pgfplotsset{every-axis label/.append style={font=\LARGE}}
\tikzset{every mark/.append style={font=\huge}}
\begin{tikzpicture}[baseline,scale=0.45]
			\begin{axis}[	
			legend style={
				at={(0,0)},
				anchor=north east},	
			align =center, 
			font=\LARGE,			
			title= {Accidents},
			ylabel=accuracy,
			xlabel=re-orderings
			]
			\pgfplotstableread{accidents-clustering}\table 
			\addplot[mark=+] table[x index=0,y index=1] from \table;
			\pgfplotstableread{accidents-asclustering}\table 
			\addplot[mark=o] table[x index=0,y index=1] from \table;
			\pgfplotstableread{accidents-redudancy}\table 
			\addplot[mark=*] table[x index=0,y index=1] from \table;
						\addlegendentry{ $\mathtt{bipDCC}$}
			\addlegendentry{ $\mathtt{bipACC}$}
			\addlegendentry{ $\mathtt{bipLC}$}
\end{axis}
\end{tikzpicture}
\begin{tikzpicture}[baseline,scale=0.45]
			\begin{axis}[
			align =center, 
			font=\LARGE,			
			title= {Earthquakes},
			ylabel=accuracy,
			xlabel=re-orderings
			]
			\pgfplotstableread{earthquake-clustering}\table 
			\addplot[mark=+] table[x index=0,y index=1] from \table;
			\pgfplotstableread{earthquake-asclustering}\table 
			\addplot[mark=o] table[x index=0,y index=1] from \table;
			\pgfplotstableread{earthquake-redundancy}\table 
			\addplot[mark=*] table[x index=0,y index=1] from \table;
\end{axis}
\end{tikzpicture}
%
\caption{Coherence accuracy (vertical axis) vs. number of sentence re-orderings (horizontal axis) per document, for our  $\mathtt{bipDCC}$ (+),  $\mathtt{bipACC}$ (o) and  $\mathtt{bipLC}$ ($\bullet$) metrics, for \textit{Accidents, Earthquakes}.}
\label{fig:perm}
\end{figure}

Figure \ref{fig:perm} shows the average accuracy of the $n^{th}$ re-ordered document relative to the original document for each of the coherence metrics, separately for the Accidents and Earthquakes subsets. A perfect coherence metric would be a straight line with an accuracy of $1$ as the coherence score of the original document would \emph{always} be larger than a re-ordered version. Instead, we see fluctuations that diverge more (for $\mathtt{bipDCC}$ and $\mathtt{bipACC}$) or less ($\mathtt{bipLC}$) from that ideal straight line. Consistently with Table \ref{tab:coh}, $\mathtt{bipLC}$ has the most accurate and most robust performance. Of interest are the extreme peaks and drops as $n$ increases: whereas for Accidents accuracy plummets for $\mathtt{bipDCC}$ and $\mathtt{bipACC}$ , for Earthquakes accuracy shoots up for $\mathtt{bipDCC}$ and $\mathtt{bipACC}$ but drops for $\mathtt{bipLC}$. Closer inspection reveals that these more or less dramatic fluctuations are likely due to data sparsity: many documents are shorter than 40 sentences; thus, for high $n$ values, there are fewer documents where it is possible to do $n$ permutations, and consequently only a few documents determine the accuracy, making the overall findings less generalisable. 

\section{Retrieval Evaluation}
\label{s:results-ir}
We now test the usefulness of our coherence metrics to retrieval. 

\subsection{Integration of Coherence to Ranking}
\label{ss:integration}
Our assumption is that more coherent documents are likely to be more relevant. 
To test this we rerank the top 
1000\footnote{\scriptsize{Limiting the reranking to the top 1000 is more efficient than reranking all documents with a nonzero baseline score, without making a large difference to system effectiveness \cite{Cras05}.} } documents retrieved by a baseline model according to their coherence scores. 
In doing so, we treat document coherence as a type of query independent aspect of document quality that we combine with a query dependent baseline. The main idea is: (i) attach a static weight to each document based on its coherence; and (ii) combine this weight with the query dependent baseline score, to give a new score and ranking.  For step (ii) we choose to use three types of linear combination which make it intuitively easy to interpret the impact of the coherence score on the final ranking. We present these next.

Let $B$ be the baseline ranking score of a document. 
Let $C$ be the coherence score of a document, computed according to each of our three coherence metrics presented in Section \ref{s:model}. 
Let $R$ be the reranking score of a document, which should combine both $B$ and $C$. We compute  $R$ as: $ R =  B + \widehat{C}$,
 where $\widehat{C}$ is a transformation of the document coherence score. 
We transform this document coherence score, in three different and increasingly parameterised ways, known to smooth out the integration of document quality features in general into ranking. Specifically we use the \textit{log}, \textit{satu} and \textit{sigmoid} transformations \cite{Cras05}, shown below:

\begin{equation}
\label{eq:satu}
\textrm{log}(C,w) = w \log{C}
\end{equation}
\noindent where $w$ is a smoothing parameter. 

\begin{equation}
\label{eq:satu}
\textrm{satu}(C,w,k) = w \frac{C}{k+C} 
\end{equation}
\noindent where $w$ approaches the maximum as $C$ increases, and $k$ is a parameter controlling the value of $C$. The function \textit{satu} can be reformulated as a \textit{sigmoid} by introducing another parameter:

\begin{equation}
\label{eq:sigmoid}
\textrm{sigmoid}(C,w,k,\alpha) = w \frac{C^{\alpha}}{k^{\alpha}+C^{\alpha}} 
\end{equation}
\noindent where $\alpha$ is an extra parameter allowing for more fine smoothing. See \cite{Cras05} for a discussion of the rationale and behaviour of the \textit{log}, \textit{satu} and \textit{sigmoid} transformations.

\begin{table*}[]
\centering
\scalebox{0.9}{
\begin{tabular}{lcccccc}\hline
			Method & MRR   & P@10  & ERR@20 &NDCG@20 &MAP@1000\\
			\hline
LM (baseline)			           	& 46.08 	& 31.19 	& 15.78 	& 15.68	&09.67\\
LM $\odot$ $\mathtt{bipDCC}$			&\cellcolor{gray!10}49.09	&\cellcolor{gray!10}34.00	&\cellcolor{gray!10}16.89	&\cellcolor{gray!10}16.53	&\cellcolor{gray!10}10.15\\
LM $\oplus$ $\mathtt{bipDCC}$ 		&\cellcolor{gray!10} 48.62 	& \cellcolor{gray!10}34.00 &\cellcolor{gray!10} 18.63 &\cellcolor{gray!10} 16.61	&\cellcolor{gray!10}10.07\\
LM $\otimes$ $\mathtt{bipDCC}$ 		&\cellcolor{gray!10} 47.66 	& \cellcolor{gray!10}33.60  &\cellcolor{gray!10} 18.38 &\cellcolor{gray!10}16.26 	&\cellcolor{gray!10}10.01\\
LM $\odot$ $\mathtt{bipACC}$		&\cellcolor{gray!10}49.14	&\cellcolor{gray!10}34.20	&\cellcolor{gray!10}16.94	&\cellcolor{gray!10}16.60	&\cellcolor{gray!10}\bf10.18\\
LM $\oplus$ $\mathtt{bipACC}$ 		&\cellcolor{gray!10} 48.89 	&\cellcolor{gray!10} \bf34.40 &\cellcolor{gray!10} 18.11 &\cellcolor{gray!10} 16.52 	&\cellcolor{gray!10}10.09\\
LM $\otimes$ $\mathtt{bipACC}$ 		&\cellcolor{gray!10} 47.76 	&\cellcolor{gray!10} 33.20 & \cellcolor{gray!10}17.02 &\cellcolor{gray!10} 16.20 	&\cellcolor{gray!10}10.01\\
LM $\odot$ $\mathtt{bipLC}$			&\cellcolor{gray!10} \bf52.82	&\cellcolor{gray!10}32.80	&\cellcolor{gray!10} \bf18.69	&\cellcolor{gray!10}16.67	&\cellcolor{gray!10}10.05\\
LM $\oplus$ $\mathtt{bipLC}$ 		&\cellcolor{gray!10} 47.63 	&\cellcolor{gray!10} 33.41 &\cellcolor{gray!10} 16.50 &\cellcolor{gray!10} \bf16.76	&\cellcolor{gray!10}10.12\\
LM $\otimes$ $\mathtt{bipLC}$ 		&\cellcolor{gray!10}50.52 &\cellcolor{gray!10} 32.60  &\cellcolor{gray!10} 17.28 &\cellcolor{gray!10} 16.48	&\cellcolor{gray!10}09.92\\
\hline
\end{tabular}
}
\caption{Retrieval performance of coherence-based reranking. Improvements over the baseline are shaded and single best scores per evaluation measure are in bold.}
\label{tab:res}
\end{table*}

\subsection{Experimental Setup}
We compare retrieval performance between
\begin{enumerate}
\item a baseline ranking model (query likelihood language model with Dirichlet-smoothing, denoted LM) that does not use coherence, and 
\item nine reranked versions of that baseline ranking that use document coherence (the three coherence metrics $\mathtt{bipDCC}$, $\mathtt{bipAcc}$, $\mathtt{bipLC}$ presented in Section \ref{s:model} combined with the three integrations to ranking ($\textit{log}$ denoted $\odot$, $\textit{satu}$ denoted $\oplus$, $\textit{sigmoid}$ denoted $\otimes$) presented in Section \ref{ss:integration}).
\end{enumerate} 
We retrieve documents from the ClueWeb09 cat. B dataset using queries 150-200 from the Web AdHoc track of TREC
2012.   
We use the Indri IR system without stemming and without removing stopwords. 
Following \cite{petersen}, we remove spam from ClueWeb09 cat. B using the spam rankings of Cormack et al.
\shortcite{fusion} 
with a percentile-score $<90$. This is a much higher threshold than the $<70$ recommended in \shortcite{fusion}, practically meaning that we use much stricter spam filtering than recommended.   
We evaluate retrieval at different rank positions with MRR, Precision@10 (P@10), ERR@20, NDCG@20, and MAP@1000. 

The baseline and our reranking methods include parameters $\mu$ (for Dirichlet smoothing), and $w,k,\alpha$ that we tune using 5-fold cross-validation. We report the average of the five test folds. 
We vary $\mu \in$ [100, 500, 800, 1000, 2000, 3000, 4000, 5000, 8000, 10000], $w \in [0.0,2.0]$ in steps of $0.1$, $k \in [0.0,2.0]$ in steps of $0.1$, and $\alpha \in [0.0,1.0]$ in steps of $0.1$.

\subsection{Retrieval Findings}

\subsubsection{Retrieval Effectiveness}

Table \ref{tab:res} displays the retrieval effectiveness of our coherence-based reranking experiments and the original ranking baseline. Coherence-based reranking improves over the baseline at all times. The best overall performance differs per evaluation measure: for MRR, ERR@20 and NDCG@20 our strongest coherence metric (as shown in the previous section), $\mathtt{bipLC}$, is the best; for P@10 and MAP, $\mathtt{bipACC}$ is the best. All three MRR, ERR@20 and NDCG@20 are evaluation metrics of early precision: MRR measures the rank of the first relevant document, while ERR@20 and NDCG@20 focus in the top 20 ranks (they both consider the rank of a document, but they differ in that ERR conditions the usefulness of a document at rank $i$ on the usefulness of the documents at ranks less than $i$, whereas NDCG assumes the usefulness of a document to be independent of the documents ranked above it). So it seems that $\mathtt{bipLC}$ is best for early precision measures. However, $\mathtt{bipACC}$ is best for P@10 (precision in the top 10 ranks), which is also an early precision measure. This could be due to the way P@10 computes precision, namely as the number of relevant documents in the top 10 but regardless of their ranking. In effect this transforms the top 10 into an unordered set of documents, whose measurement is not guaranteed to agree with rank order-oriented measures such as ERR and NDCG. 

We also see that the $\mathtt{bipDCC}$ coherence metric is never the best. This could be because $\mathtt{bipDCC}$ does not account for the number of entities that are shared by sentences. As this is a major indication of coherence (topic transition across sentences), it is likely that failing to account for this degrades coherence prediction (as we also saw in Table \ref{tab:coh} on average and for the Accidents dataset). As a result, using for reranking a weaker coherence metric ($\mathtt{bipDCC}$) improves retrieval less than when using our other two stronger coherence metrics. Note however that even though $\mathtt{bipDCC}$ is not the strongest coherence metric, it still benefits retrieval performance compared to the baseline.

Overall the difference in performance among the coherence runs in Table \ref{tab:res} is relatively small, except for MRR. The MRR exception is because the MRR score tends to change substantially for differences in even one rank position. For instance, when the first relevant document is at rank 1, MRR = 100; when at rank 2, MRR = 50.00; when at rank 3, MRR = 33.33, and so on. Considering this, even the largest difference in MRR among our coherence runs (from 52.82 to 47.63) is not indicative of considerable variation in rank position.  

We also see in Table \ref{tab:res} that even though $\textit{sigmoid}$ ($\otimes$) is more parameterised than the other two combinations, it is never the best. Instead, $\textit{log}$ ($\odot$) and $\textit{satu}$ ($\oplus$) take turns at being best, indicating that the coherence-based reranking performance is not a byproduct of additional tuning parameters that smooth out retrieval regardless of coherence. 

We further note that improvements over the baseline for MAP@1000 are smaller than improvements over the baseline for the other early precision measures. This is not surprising: typically as the depth of the measured precision increases, for instance from ranks 10--20 to rank positions $>$500, the actual precision score averaged over all retrieved documents up to that rank progressively deteriorates, because increasingly less relevant documents enter the ranking. 

Finally, to contextualise the performance of our coherence metrics, we report that the retrieval performance of the two best \textit{non-bipartite} coherence metrics in Table \ref{tab:coh}, out-degree and entity distance, never exceeds the scores of our best bipartite metrics\footnote{\scriptsize{The respective maximum scores of either out-degree or entity distance are MRR: 34.18, P@10: 22.40, ERR@20: 15.86, NDCG@20: 14.66, and MAP@1000: 07.22.}}.

\subsubsection{Coherence and Query Difficulty}
An aggregated overview of if and how much coherence-based reranking improves performance for queries of various levels of difficulty can be seen in Table \ref{tab:query}. The percentages in Table \ref{tab:query} have been produced as follows. For each retrieval precision measure, we rank all queries decreasingly according to their baseline retrieval score. We then use these scores to sort queries into the four quantiles (Q1--Q4) shown in columns 2--5. In effect these quantiles group queries according to their difficulty; Q1 contains those queries that have the highest baseline retrieval score (hence they are perceived as easier queries for the IR system to satisfy), whereas Q4 contains those queries with the lowest baseline retrieval score (which are perceived as the hardest for the IR system to satisfy). For the queries in each quantile, we compute their absolute difference in score between the baseline and each coherence-based reranking and turn this difference into a percentage. The percentages of each quantile correspond to the average improvement in retrieval performance over the baseline per quantile. This is the average over all queries in the quantile, and over all retrieval measures.  

We see that Q4 gains the most, on average across all coherence runs. As Q4 corresponds to the hardest queries that an IR system has to process, this means that reranking by coherence can improve performance for those queries that standard ranking has the most trouble with. However, the percentages per coherence metric show that our strongest coherence metric, $\mathtt{bipLC}$, benefits mostly Q2 -- Q3, and very little or not at all Q4. This very small or no improvement in Q4 for $\mathtt{bipLC}$ is in fact an artefact of how we computed the percentages: because it is not possible to compute a percentage improvement over zero, we removed from Q4 those queries that had a zero baseline score. These were on average 10.5 queries per evaluation measure. Removing these highly difficult queries has the effect of underestimating the impact of coherence-based reranking in particular in Q4, and especially for $\mathtt{bipLC}$.

Overall, the smallest improvement for all coherence-based reranking is in Q1, which corresponds to queries that baseline IR ranking can cope with satisfactorily. This indicates that the margin for improvement over the baseline may be smaller for those queries. 

\begin{table}[]
\centering
\scalebox{0.8}{
\begin{tabular}{lrrrr}\hline\noalign{\smallskip}
Method 		&Q1	&Q2	&Q3	&Q4\\
\hline
LM $\odot$ $\mathtt{bipDCC}$     &+4\%&\cellcolor{blue!25}+12\%&+2\%&\cellcolor{blue!10}+9\%\\
LM $\oplus$ $\mathtt{bipDCC}$   &+2\%&\cellcolor{blue!10}+11\%&+8\%&\cellcolor{blue!25}+219\%\\
LM $\otimes$ $\mathtt{bipDCC}$     &+3\%&\cellcolor{blue!10}+12\%&0\%&\cellcolor{blue!25}+140\%\\
LM $\odot$ $\mathtt{bipACC}$    &+3\%&\cellcolor{blue!25}+16\%&+3\%&\cellcolor{blue!10}+10\%\\
LM $\oplus$ $\mathtt{bipACC}$  &+1\%&\cellcolor{blue!10}+21\%&+4\%&\cellcolor{blue!25}+86\%\\  
LM $\otimes$ $\mathtt{bipACC}$    &+3\%&\cellcolor{blue!10}+10\%&+2\%&\cellcolor{blue!25}+23\%\\
LM $\odot$ $\mathtt{bipLC}$   &0\%&\cellcolor{blue!25}+45\%&\cellcolor{blue!10}+19\%&--2\%\\
LM $\oplus$ $\mathtt{bipLC}$   &+3\%&\cellcolor{blue!10}+18\%&\cellcolor{blue!25}+30\%&+3\%\\
LM $\otimes$ $\mathtt{bipLC}$   &--2\%&\cellcolor{blue!10}+24\%&\cellcolor{blue!25}+28\%&+7\%\\
\hline
Average	&+2\%&\cellcolor{blue!10}+19\%&+11\%&\cellcolor{blue!25}+55\%\\
\hline
\end{tabular}
}
\caption{Average improvement in retrieval performance over the baseline. Darker cells mark higher improvements.}
\label{tab:query}
\end{table}

\subsubsection{Error Analysis} 

To gain more insight into the type of contribution that coherence makes to retrieval, we look at those cases that benefit the most from coherence-based reranking. 
For query 174 (\texttt{rock art}), the documents ranked in the top two places by the baseline retrieval model receive a coherence score of 0.0 by all of our coherence metrics. These top two documents have no TREC relevance assessments (hence most IR evaluation metrics will treat them as non-relevant). Even though these documents have stayed in the dataset after we filtered out spam (using very strict spam thresholding, as discussed above), manual inspection reveals these documents to be largely non-informative. The first few lines of these documents are included below:
\begin{quote}
\scriptsize
[clueweb09-en0009-40-30672]: Music Democracy :: Unchain You Art username: Damn! I forgot my password password: The Music Democracy Team is attending MIDEM 09 in Cannes. If you're interested in a meeting, please contact us (link at the bottom of the page) Tests are underway and certain features could be unavailable punctually.We apologize for these inconveniences. HOME URBAN ROCK ELECTRO POP BLUES WORLD VARIOUS Registration as Musician * Username  * Email * Retype email * Password (at least 6 characters) * Retype Password * Country (included province) Select Albania Algeria American Samoa Andorra Angola Anguilla Antarctica Antigua and Barbuda Argentina Armenia Aruba [$\cdots$]

[clueweb09-en0000-95-09794]: Outline of Art History - Ancient Art Search Art History Home Education Art History Email Art History Artists Styles Works of Art Filed In: Art History Outline of Art History - Ancient Art 30,000 BC - c. 400 AD Outline of Art History Part 1: Ancient Art Part 2: Medieval Art Part 3: Renaissance Art Part 4: Modern Art Part 5: Contemporary Art Related Resources Ancient Art Resources Prehistory Paleolithic (Old Stone Age) [$\cdots$]
\end{quote}

For documents like these, the baseline ranking function (which considers solely single term frequencies) has no way of detecting low document informativeness. The extremely frequent and uninformative (almost spam-like) repetition of the same terms, not only goes undetected in the baseline ranking, but can also result in the documents being ranked very high when they contain query terms (this is what happened for query 174). Our coherence metrics are particularly useful in these cases, because they can detect the low quality of these documents. 

Of interest is also the document ranked by the baseline in position 4 for the same query 174. This document also has no TREC relevance assessments, and receives the following coherence scores: $\mathtt{bipDCC}$=0.009, $\mathtt{bipACC}$=0.022, $\mathtt{bipLC}$=0.0. This document is a wikipedia listing of museums in Maryland:
\begin{quote}
\scriptsize
[clueweb09-enwp01-26-04667]: List of museums in Maryland [$\cdots$] encompasses museums, defined for this context as institutions (including nonprofit organizations, government entities, and private businesses) that collect and care for objects of cultural, artistic, scientific, or historical interest and make their collections or related exhibits available for public viewing. Museums that exist only in cyberspace (i.e., virtual museums) are not included. Lists of Maryland institutions which are not museums are noted in the "See also" section, below. To use the sortable table, click on the icons at the top of each column to sort that column in alphabetical order; click again for reverse alphabetical order. Name Location Region Area of study Summary Aberdeen Room Archives \& Museum Aberdeen Local history website Academy Art Museum Easton Art website, works on paper and contemporary works by American and European masters Adkins Historical Museum Mardela Springs Open air website, eight historic buildings and the gravestones of a Revolutionary War patriot and his wife, buildings open by appointment African-American Heritage Society Museum [$\cdots$]
\end{quote}

Both our clustering-based coherence metrics ($\mathtt{bipDCC}$ and $\mathtt{bipACC}$) give to this document weights that concentrate on the dense clusters of entities. On the contrary, the $\mathtt{bipLC}$ metric emphasises more the transition of entities across sentences (which is extremely low in this document, as new entities (museum names and themes) keep on being introduced, mentioned in 1-2 sentences, and then quickly dropped). This is a specific discourse feature of text (to list or enumerate themes without linking them into the discourse), which goes undetected by clustering ($\mathtt{bipDCC}$ and $\mathtt{bipACC}$), but not by $\mathtt{bipLC}$. Note that this document was ranked as the fourth most relevant document for the query \texttt{rock art}. After manual inspection, we consider it neither very relevant, nor very coherent. 

The above are examples of relatively low quality documents that are ranked (erroneously) high by the baseline but receive a very low coherence score (correctly) by our coherence metrics. Next we display examples of the opposite: high quality documents that are ranked (correctly) higher by coherence-based reranking than by the baseline.

\begin{quote}
\scriptsize
[clueweb09-enwp00-52-08632]: Rock art of the Chumash people. [$\cdots$]
 Chumash Rock Art is a type of artwork created by the Chumash people, mainly in caves or on cliffs in the mountains in areas of southern California. Contents: Chumash people, Rock Art Locations, Shamans and Visions, Shamans and Rock Art, Rock Art Characteristics, Meanings of Rock Art, Conclusion, References
 [$\cdots$] Chumash Rock Art is almost invariably found in caves or on cliffs in the
mountains, although some small, portable painted rocks have been discovegray by
Campbell Grant. The rock art sites are always found near streams, springs, or
some other source of permanent water. In his research of southern California
rock art, Grant recorded numerous sites from different areas that were all
close to a water source. He found twelve painted sites in the highest parts of
the mountainous Chumash territory, the Ventureno area  [$\cdots$]
\end{quote}

\begin{quote}
\scriptsize
[clueweb09-en0009-97-31173]: The Heilbrunn Timeline of Art History. The Metropolitan Museum of Art. African Rock Art.  African Rock Art.
 Thematic Essay Categories.
 Recent Additions.
 All Thematic Essays. African Art. Central Africa. Eastern Africa. Southern Africa. Western Africa
[$\cdots$] Africa's oldest continuously practiced art
form. Depictions of elegant human figures, richly hued animals, and figures
combining human and animal features called the rianthropes and associated with
shamanism continue to inspire admiration for their sophistication, energy, and
direct, powerful forms. The apparent universality of these images is deceptive;
content and style range widely over the African continent. Nevertheless,
African rock art can be divided into three broad geographical
zones southern, central, and northern. The art of each of these zones is
distinctive and easily recognizable, even to an untrained eye [$\cdots$]
\end{quote}

The above documents are \textit{both} coherent \textit{and} relevant to query 174. Coherence-based reranking moves these documents three and two positions higher up in the ranking, respectively. Interestingly, for both of these documents $\mathtt{bipACC}$ gives the highest coherence score. This is because both these documents first list an index of subtopics that they contain and then discuss each of them under different specialised subheadings. This type of discourse, which is characterised by several local clusters that are however closely linked to each other by an underlying common theme (rock art in this case) is best modelled by $\mathtt{bipACC}$, which focuses on local clusters (unlike $\mathtt{bipLC}$) while also accounting for the number of entities that are shared by sentences (unlike $\mathtt{bipDCC}$).

\section{Discussion}
\label{s:discussion}

In Section \ref{s:model} we present three bipartite metrics for approximating document coherence. Our metrics are not the only tailor-made metrics for bipartite graphs. There is substantial work in the
field of network science aiming at defining metrics that highlight certain topological features of graphs,
several of which may be related to coherence, but to the best of our knowledge this has not been done so far. For example, several techniques exist for community detection
~\cite{fortunato}, which may potentially be used in combination with bipartite graph distance to afford
more precise coherence scores. The rationale is that, if a document is coherent, sentences being identified as belonging
to a particular community should be close in the document. Similarly, several techniques exist for detecting
maximal bicliques~\cite{biclique} (i.e., maximal subsets of vertices where all $\top$ and $\bot$-vertices are connected). Intuitively, maximal biclique detection could be used to detect document sentences and entities so related that
there is no doubt they represent a coherent flow of discourse. 

Regarding the integration of document coherence into ranking, we treat our document coherence metrics as a query-independent type of document quality score that we combine linearly with the retrieval status value of the baseline ranking to rerank retrieved documents. This is a straightforward reranking approach; more involved
reranking functions  \cite{Cras05} may be used to possibly improve retrieval effectiveness. For instance, studying the distribution of the document coherence scores as well as the distribution of the document relevance scores can inform functions that aim at fitting the former to the latter more closely. Another integration method that can be used is rank fusion. The idea here is to the turn the baseline and the coherence scores into two rankings, which are then fused, e.g. using CombMNZ, voting algorithms, or Bayesian inference \cite{beitzel2003}. This has the advantage of ignoring the coherence score distributions, so for example heavily skewed distributions of document coherence cannot be allowed to have too much impact upon the final ranking. Alternatively, document coherence can also be turned into a prior probability of relevance and combined with the language modelling baseline probability, which would produce a seamless ``coherence-enhanced language model''. All of the above directions are interesting to pursue in the future.

\section{Conclusions}
\label{s:conc}
We presented three novel bipartite graph metrics of document coherence. Our metrics extend the state of the art in unsupervised coherence modelling by approximating coherence directly on bipartite graphs of discourse entities and sentences (unlike previous methods that use their one--mode projections). We experimentally evaluated the accuracy of our metrics in modelling coherence. Our bipartite metrics incurred no additional efficiency cost over existing one--mode graph metrics. One of our metrics was found to be \textit{much} more accurate approximation of document coherence than the state of the art computed from one-mode projections \cite{strube,petersen}. We also experimentally evaluated the usefulness of our document coherence metrics to IR, and found them overall successful, and in particular for early precision and ``difficult'' queries. Our results can be seen as another piece of evidence in a long string of results showing that algorithmic approximations of document \emph{quality} can be exploited in IR to obtain better retrieval performance.



%

\bibliographystyle{abbrv}
\bibliography{bib}
\balancecolumns
\end{document}